\documentclass[]{aa}

\usepackage{graphicx}

\begin{document}

\title{Models of the 
formation of the planets in the 47 UMa system}

\author{K. Kornet\inst{1} P. Bodenheimer\inst{2} \and M. R{\'o}{\.z}yczka\inst{1}}

\offprints{K. Kornet}

\institute{Nicolaus Copernicus Astronomical Center , Bartycka 18  , Warsaw, PL-00-716, Poland\\ \email{kornet@camk.edu.pl}, \email{mnr@camk.edu.pl}
\and
UCO/Lick Observatory, Department of Astronomy and Astrophysics,
University of California, Santa Cruz,CA 95064, USA\\
\email{peter@ucolick.org}
}

\titlerunning{Formation of 47 UMa system}
\authorrunning{Kornet, Bodenheimer \& R{\'o}{\.z}yczka}

\abstract{%
Formation of planets in the 47 UMa system is followed in an evolving
protoplanetary disk composed of gas and solids. The evolution of the disk
is calculated from an early stage, when all solids, assumed to be 
high-temperature silicates,  are in the dust form, to
the stage when most solids are locked in planetesimals. The simulation of
planetary evolution starts with a solid  embryo of $\sim$1 Earth mass, and
proceeds according to the core accretion -- gas capture model. 
Orbital parameters are kept constant, and it is
assumed that the environment of each planet is not perturbed by the second
planet. It is found that conditions suitable for both planets to form
within several Myr are easily created, and maintained throughout the
formation time, in disks with $\alpha\approx0.01$. In such disks, a planet
of 2.6 Jupiter masses (the minimum for the inner planet of the 47 UMa
system) may be formed at 2.1 AU from the star in $\sim$3 Myr, while a
planet of 0.89 Jupiter masses (the minimum for the outer planet)
may be formed at 3.95  AU from the star in about the same time.  
The formation of planets is possible as a result of a significant
enhancement of the surface density of solids between 1.0 and 4.0 AU, which
results from the evolution of a disk with an initially uniform gas-to-dust
ratio of 167 and an initial radius of 40 AU.
}

\maketitle

\section{Introduction }
Planets in the mass range 0.1 to 10 Jupiter masses (M$_J$), separated
from their central stars by 0.04 to 5 AU, have been discovered around
$\sim$ 100 main-sequence stars with masses in the range 0.3 -- 1.1
M$\odot$. The general observed properties of these planets, several of
which are in fact in planetary systems, are reviewed by Marcy, et. al,
Perryman (\cite{Marcy00}), and Bodenheimer \& Lin
(\cite{Bodenheimer02}). In comparison to the solar system, many of
these systems are unusual in two respects: they contain Jupiter-mass
planets at close distances (down to 0.04 AU) from their star, and many
of them are in orbits with substantial (0.4 to 0.7) eccentricity.
However one particular system, belonging to the star 47 Ursae Majoris
(47 UMa), stands out as being more solar-system like than most of the
other extrasolar planetary systems. This paper examines a possible
mechanism for the origin of the two planets in the 47 UMa system.

The solar-type star 47 UMa has spectral type G0V, a mass of 1.03
M$_\odot$, and a metallicity [Fe/H] = $ - 0.08$.  One of the first
extrasolar planets to be discovered was a companion to 47 UMa (Butler
\& Marcy \cite{Butler}) at 2.09 AU, with a current minimum mass of
2.62 M$_J$ and an eccentricity of 0.04.  Later (Fischer et
al. \cite{Fischer02}) a second planet was discovered; current orbital
parameters (Fischer \cite{Fischer_pr}, private communication) give a
semimajor axis of 3.95 AU, a minimum mass of 0.89 M$_J$ and an
eccentricity which is not well determined but is probably less than
0.1.  The mass ratio of the two planets (0.34) is close to that of
Saturn/Jupiter (0.30), and the ratio of semimajor axes (1.90) is also
close to that of Saturn/Jupiter (1.83).  A dynamical analysis of the
system (Laughlin et al. \cite{Laughlin01}) shows that if the two
planets are in the same orbital plane, Earth-mass planets in the
habitable zone would have stable orbits. However in the presence of
the two fully formed giant planets, the formation of Earth-mass
planets in the inner regions of the system is possible only interior
to the habitable zone.

Three main theories have been proposed regarding the origin of
planetary-mass objects.  The first is dynamical fragmentation of a
rotating collapsing protostar, the mechanism thought to be responsible
for multiple stellar systems (reviewed by Bodenheimer et
al. \cite{Bodenheimer00}) and possibly the isolated planetary-mass
objects observed in the young cluster $\sigma$ Orionis (Zapatero
Osorio et al. \cite{Zapatero}).  The second is gravitational
instability in a disk (Kuiper \cite{Kuiper}; Boss \cite{Boss00}; Boss
et al. \cite{Boss02}), in which, on a few dynamical time scales, a
gravitationally bound subcondensation forms in a disk that at some
location has a Toomre $Q$ value on the order unity. The third
mechanism, known as the core accretion -- gas capture process,
involves the relatively slow gradual accretion of small condensed
particles in a disk, eventually resulting in a solid core of a few
M$_\oplus$ which is able to gravitationally capture gas from the
surrounding nebular disk (Safronov \cite{Safronov}; Perri \& Cameron \cite{Perri}; Mizuno \cite{Mizuno}). The strengths and weaknesses of the second and third processes
are reviewed by Bodenheimer \& Lin (\cite{Bodenheimer02}).

The formation of the inner giant planet in the 47 UMa system has been
studied by Bodenheimer et. al (\cite{BHL}) under the assumption that
it formed {\it in situ} by the core accretion -- gas capture process.
The evolutionary calculations they performed are based on the earlier
work by Bodenheimer \& Pollack (\cite{Bodenheimer86}), who assumed a
constant solid accretion rate for the buildup of the core, and by
Pollack et al.  (\cite{Pollack96}) who employed more detailed physics,
including a (non-constant) solid accretion rate calculated from
three-body accretion cross sections. The aim of the calculation was to
determine the disk properties needed to form the planet on a
reasonable time scale at 2.1 AU. Two calculations with variable solid
accretion rate were performed (their cases U1 and U2). The two most
important parameters in such calculations are (1) the initial surface
density of solid material $\Sigma_s$, to which the formation time is
highly sensitive, and (2) the grain opacity $\kappa_g$ in the envelope
of the protoplanet, to which the formation time is moderately
sensitive (Pollack et al. \cite{Pollack96}). In both cases presented
by Bodenheimer et al. (\cite{BHL}) the values of $\kappa_g$ were based
on interstellar grain properties as calculated, for example, by
Pollack et al. (\cite{Pollack85}). In the temperature range 100 --
1500 K, those opacities are typically in the range 1--8 cm$^2$
g$^{-1}$.  The values of $\Sigma_s$ were set to 50 and 90 g cm$^{-2}$
in the two calculations, and the formation times turned out to be 18.6
Myr and 1.9 Myr, respectively. The final solid core masses were 38 and
69 M$_\oplus$, respectively, compared to a total assumed final mass of
2.5 M$_J$ in both cases.

The lifetimes of disks around young stars, which constrain the
formation times for giant planets, fall in the range 1--8 Myr (Haisch
et al. \cite{Haisch}), with half of the disks in young clusters already gone
after times of 3--4 Myr. The mass ratio of gas to solids in a
solar--composition disk is expected to be about 200 at 2.1 AU, since
the ice component is expected to be evaporated. Thus the total surface
density of the disk, $\Sigma_{total}$, would have to be 1--2 $\times
10^4$ g cm$^{-2}$, in order for the planet to form in a reasonable
time.  If one examines the steady-state disk models of Bell et
al. (\cite{Bell}) with viscosity parameter $\alpha = 10^{-2}$, one sees that
a disk with that high a $\Sigma_{total}$ at 2.1 AU would have a
temperature of about 1000 K (consistent with the evaporation of ices)
and an accretion rate onto the star of $\dot M \approx 10^{-5}$
M$_\odot$ yr$^{-1}$.  Thus some difficulties with the model include
(1) the required disk $\dot M$ is much higher than the values of $\sim
10^{-8}$ M$_\odot$ yr$^{-1}$ in typical observed disks around young
stars (Calvet et al.  \cite{Calvet}), (2) such a disk is likely to be
gravitationally unstable at larger radii (Bell et al. \cite{Bell}) which
could result in the formation of a much more massive planet at a
distance of about 10 AU, and (3) $\Sigma_{total}$ is about 20 times
that in the `minimum mass' solar nebula (Hayashi et al.  \cite{Hayashi}).

 However a basic assumption of such disk models is that the ratio of
gas to solids is constant at each radius as the disk evolves in time,
except as modified by evaporation or condensation of solids. Here we
consider alternate disk models, in which the evolution of
$\Sigma_{total}$ and $\Sigma_s$ is not necessarily coupled. It has
long been recognized that the solid particles in a disk evolve
differently than the gas; for a review of the physical processes
involved see Weidenschilling \& Cuzzi
(\cite{Weidenschilling}). However global disk models in which the
evolution of the solids and of the gas were followed consistently over
timescales comparable to disk lifetimes (a few Myr) did not become
available until Stepinski \& Valageas (\cite{SV96}, \cite{SV97})
published some numerical solutions, based on a number of
approximations in the physics.  The typical result of such
calculations was a decoupling of the evolution of the solid component
from that of the gas, once the particle size had become large
enough. At the end of a simulation, a typical disk had a region, say
from 1 to 10 AU, where the ratio of surface densities of solid and gas
was considerably higher than that suggested by solar composition, and
also regions where solids were practically absent. For some disk
models, however, all of the solid material accreted onto the star.
Results for a larger regime of parameter space for disk models were
presented by Kornet et. al (\cite{Kornet}), based on the methods of
Stepinski \& Valageas (\cite{SV96}, \cite{SV97}) but with further
simplifications. These models start with a uniform, solar ratio of
solids to gas, and evolve for $10^7$ yr, following the buildup of the
initially small particles up to the size range 1--10 km.  The final
stages of planet formation are not considered; a wide variety of
possible distributions of solid material in evolved disks is found.
The goal of the present paper is to investigate whether the formation,
on a time scale of a few Myr, of both planets in the 47 UMa system can
be explained with reasonable disk models, based on the calculations of
Kornet et al. (\cite{Kornet}).

\section{Method of calculation}
\subsection{Disk models}

The method of calculation is described by Kornet et al.
(\cite{Kornet}). The gas component is modeled in one space dimension
by an analytic solution to the viscous diffusion equation, which gives
the surface density of the gas as a function of radius $r$ and time
$t$ (Stepinski \cite{Stepinski}).  The viscosity is given by the usual
$\alpha$ model. The temperature of the gas is calculated in the
thin-disk approximation, assuming vertical thermal balance, according
to equations (2) through (6) in Stepinski (\cite{Stepinski}).
 
The main assumptions used in the calculation of the evolution of the
solid component are (1) at each radius the particles are all assumed
to have the same size, (2) there is only one component of dust, in
this case corresponding to high-temperature silicates, which have an
evaporation temperature of 1350 K, (3) all collisions between
particles lead to coagulation, (4) when the temperature exceeds the
evaporation temperature, the solids are assumed to be in the form of
vapor which evolves at the same radial velocity as the gas component,
(5) when the disk temperature falls below the evaporation temperature
at a given radius, all of the local vapor is assumed to condense
immediately into grains with particle size 10$^{-3}$ cm, (6) the
radial velocity of solid particles is determined by the effects of gas
drag.  The vertical thickness of the solid particle
distribution at each radius is calculated and is evolved in time, so
the effect of sedimentation of grains toward the midplane is taken
into account.  The evolution of solids does not affect the evolution
of the density or the temperature of the gas.

The equations solved for the evolution of the solids include the
continuity equation, the gas drag effect, coagulation and evaporation of
particles. For the calculation of relative velocities at which
coagulation proceeds, a
turbulent model is assumed as described by Stepinski \& Valageas
(\cite{SV97}).  Those equations are solved numerically on a moving
grid. Its outer boundary follows the motion of the outer edge of the
solid disk. The ratio of radii at the inner and outer edges of the
grid is kept constant. This ratio is chosen to be small enough for the
dust velocities relative to the grid at the grid inner edge to be
negative. In this way a free outflow boundary condition can be applied
there.  At the outer edge, since the dust velocity and the grid
velocity are equal, no boundary condition is required.  The grid
points are equally spaced in log radius and their number is equal to
50 for every 3 orders of magnitude of the ratio of inner and outer
radii.  Further details of the code are given in Kornet et
al. (\cite{Kornet}).

The initial 
conditions can be parameterized by the quantities $m_0$ (the mass of the
disk in M$_\odot$), and $j_0$ (the total angular momentum of the disk
in units of  $10^{52}$ g cm$^2$ s$^{-1}$).  Once  those parameters
are chosen, the 
analytic solution of Stepinski (1998) gives the gas surface density as
a function of radius at $t = 0$. The ratio of the solid surface density 
to the gas surface density is initially set at the constant value of 
$6 \times 10^{-3} $, and the particle size is everywhere set to $10^{-3}$
cm. 
 
\subsection{Planet models}

The protoplanet consists of a solid core with a constant density of 3
g cm$^{-3}$, appropriate for high-temperature silicates, and a gaseous
envelope, both of which accrete mass according to the computational
procedures described by Pollack et al. (\cite{Pollack96}) and
Bodenheimer et al. (\cite{BHL}); however certain simplifications are
made.  The basic assumptions are: (1) the protoplanet is surrounded by
a disk with an initially uniform surface density $\Sigma_{init,s}$ of
solid material, in the form of planetesimals. All
planetesimals have the same size of 2 km (see below). The solid
surface density $\Sigma_s$ decreases with time as material accretes
onto the protoplanet.  (2) The protoplanet is assumed to be the
dominant mass in the region of its feeding zone; accretion of solids
onto other planetary embryos is not considered. Random
velocities of the planetesimals are determined by only one planet
in the feeding zone; thus they are expected to be
small. The feeding zone is
assumed to extend to 4 Hill sphere radii on either side of the
protoplanet (Kary \& Lissauer \cite{Kary}). (3) Disk--planet
interactions and the resulting torques which could cause migration of
the protoplanet through the disk are not considered. Planetesimals are
assumed to be well mixed through the feeding zone at each time; thus
the value of $\Sigma_s$ is always uniform in space but usually
decreasing with time. Planetesimals do not migrate into the feeding
zone from outside, or vice versa. (4) Orbital parameters are kept
constant, and it is assumed that the environment of each planet is not
perturbed by the second planet.

Under these assumptions, the rate of accretion of solid
material onto the protoplanet, taking into account the physical cross
section of the growing planet as well as the gravitational enhancement
factor, is given by the standard expression
\begin{equation}
 { \dot M_Z } =  {\pi R_c^2 \Sigma_s \Omega F_g} 
\label{mdot1}
\end{equation}
where $\Omega$ is the orbital frequency, $R_c$ is the effective 
capture radius of the protoplanet, and $F_g$ is the gravitational
enhancement factor. To simplify the calculation of  $F_g$  we modify this formula and use an
expression given by  Papaloizou \&  Terquem (\cite{Papaloizou}): 
\begin{equation}
 {\dot M_Z} = C_1 \pi R_c R_H \Sigma_s \Omega 
\label{mdot2}
\end{equation}
 where
$R_H$ is the Hill sphere radius. The value of $C_1$ given by
Papaloizou \& Terquem (\cite{Papaloizou}) is 81/32; we use a factor of
5. An expression of the form (2) has been shown to be
consistent with the one of form (1) by Papaloizou \& Terquem
(\cite{Papaloizou}). Accretion rates from equation (\ref{mdot2}) are generally a
factor 3 lower than those obtained from a combination of equation (\ref{mdot1})
and the calculations of $F_g$ given by Greenzweig \& Lissauer
(\cite{Greenzweig}), which were used in the calculations of Pollack et
al. (\cite{Pollack96}) and Bodenheimer et al. {\cite{BHL}).

The calculation of $R_c$, the effective capture radius, takes into
account the capture of planetesimals in the gaseous envelope. The
procedure for taking into account the interaction of planetesimals
with the envelope is described by Pollack et al. (\cite{Pollack96}),
based on the work by Podolak et al. (\cite{Podolak88}). In the present
calculations an approximate fit is made to the results of Bodenheimer
et al. (\cite{BHL}), provided by Hubickyj (\cite{Hubickyj01}).
It was shown by Pollack et al (\cite{Pollack96}) that at least
for the case of Jupiter forming at 5 AU, the total formation time is
insensitive to the planetesimal size assumed, in the range 1--100 km.
For core masses less than 5 M$_\oplus$ the value of $R_c$ is simply
the core radius; for larger core masses the ratio of $R_c$ to the core
radius increases to about a factor 5.

The structure of the gaseous envelope is determined from the
equations of mass conservation, hydrostatic equilibrium, energy
generation from accretion of planetesimals and quasi-static
contraction, and radiative or convective
energy transport, as given in Bodenheimer \& Pollack
(\cite{Bodenheimer86}). To avoid excessively large temperature gradients
(which induce numerical instabilities), the energy deposition arising
from the planetesimals landing on the core is smoothed over a region
of about one core radius in extent.  The molecular opacity in the
envelope is based on calculations by Alexander \& Ferguson
(\cite{Alexander}). The grain opacity at temperatures less than the
evaporation temperature of the most refractory species, taken to be
1800 K, is set at a constant value of 0.03 cm$^2$ g$^{-1}$.  This
value is a factor of 50--100 less than the opacities obtained for
grains with interstellar properties (Pollack et
al. \cite{Pollack94}). The coagulation and settling of grains in the
atmosphere of a protoplanet results in a substantial reduction of
opacity as compared with the interstellar values; a preliminary
calculation by Podolak (2002) shows that in one particular case the
maximum grain opacity in the radiative region of a protoplanet is only
0.02 cm$^2$ g$^{-1}$. Thus the value assumed above may be considered
to be a conservative upper limit and would tend to overestimate the
formation time.  It is known that the formation time of a planet
decreases as the opacity is reduced; for example Hubickyj et
al. (\cite{Hubickyj02}) show that for a standard Jupiter model forming
at 5 AU from the Sun, a reduction of a factor 50 in the grain opacity
results in a reduction in the formation time by a factor of 2.2.  The
equation of state is non-ideal in the interior of the envelope; the
tables of Saumon et al. (\cite{Saumon}), are used, interpolated to a
near-protosolar composition of $X = 0.74, Y = 0.243, Z = 0.017$.
\begin{figure}
\resizebox{\hsize}{!}{\includegraphics{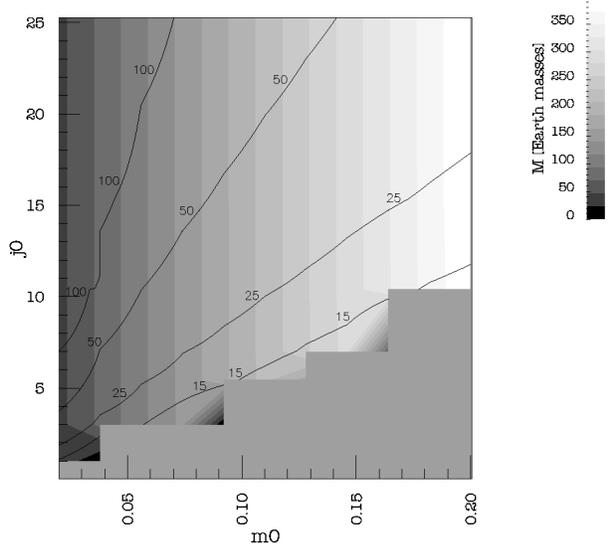}}
\caption{Final mass and outer radius of the solid disk, as 
functions of the initial disk mass $m_0$ (in solar masses)  and
angular momentum $j_0$ (in units of $10^{52}$ g cm$^2$ s$^{-1}$).
The contours give the outer radius in AU, and the grey scale
gives the mass in M$_\oplus$. The grey  region at the lower right 
indicates disks in which the solid component has completely
accreted onto the star.}
\label{mass_rout}
\end{figure}
 
Boundary conditions at the inner edge of the envelope set the
luminosity $L_r = 0 $ and the radius $r = R_{core}$, where $R_{core}$
is determined from the current core mass and the core density.  At the
outer edge temperature and density are given by the disk conditions at
the appropriate distance from the star.  The outer radius of the
planet is assumed to fall at a modified accretion radius $R_a$. Let
the tidal, or Hill, radius be
\begin{equation}
{R_H} = {a}{\biggl( { {M_p} \over {3 M_\star}}\biggr)^{1/3}}
\end{equation}
where $a$ is the distance to the central star, $M_p$ is 
the planet's mass,  and $M_\star$ is the star's mass.
Then $R_a$ is given by 
\begin{equation}
{R_a} = { { GM_p}\over {{c^2} + {{GM_p} \over {R_H}}}}
\end{equation}
where $c$ is the sound speed in the nebula.
In the limits of large and small $R_H$, this expression reduces to the 
accretion radius and the tidal radius, respectively. 
The gas accretion rate is determined by the requirement that the outer
radius of the protoplanet be close to $R_a$, within a small tolerance.       
At every time step mass is added at the outer edge so that this 
requirement is satisfied.  

\begin{figure}
\resizebox{\hsize}{!}{\includegraphics{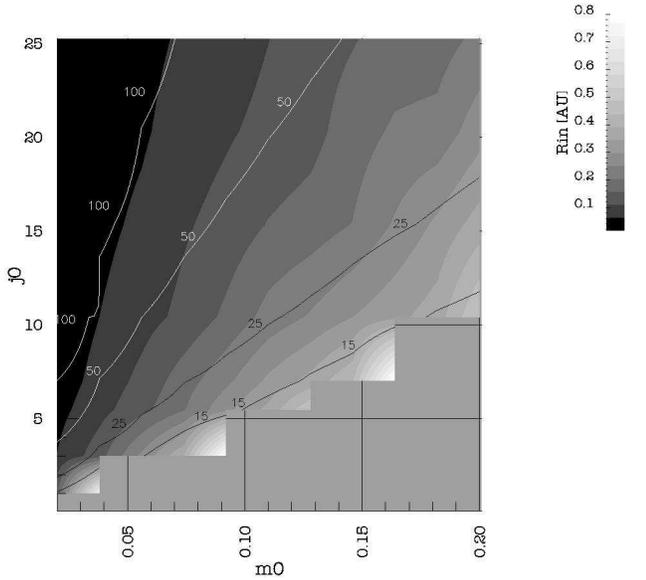}}
\caption{Final inner radius and outer radius of the solid
disk, as functions of the initial disk mass $m_0$ (in solar units) and
angular momentum $j_0$ (in units of $10^{52}$ g cm$^2$ s$^{-1}$).  The
contours give the outer radius in AU, and the grey scale gives the
inner radius in AU.  The grey region at the lower right indicates
disks in which the solid component has completely accreted onto the
star.}
\label{rin_rout}
\end{figure}

The limiting gas accretion rate onto the planet is determined by the
rate at which the nebula it is able to  supply gas.  Following Bodenheimer 
et al. (2000), we adopt for the latter a value of $3 \times 10^{-8}$
M$_\odot$/yr or $\approx 10^{-2}$ M$_\oplus$/yr, typical for the
observed protoplanetary disks.   Calculations are
generally carried to the point where the limiting rate is reached. By
that time the envelope mass has exceeded the core mass and the planet
rapidly accretes gas up to its final mass, with only a relatively
small change in the core mass.  The formation time is in effect
determined by the time needed to reach the crossover mass (envelope
mass = core mass), so the calculation is stopped just beyond that
point.

\section{Results}
\begin{figure}
\resizebox{\hsize}{!}{\includegraphics{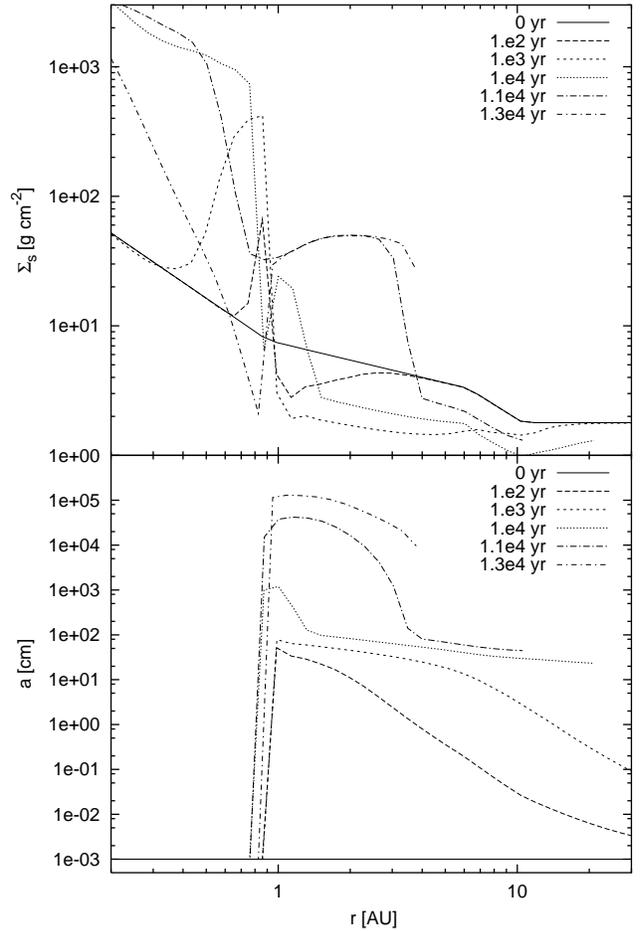}}
\caption{Evolution of the solid component in the disk model
used for planet formation in the 47 UMa system ($m_0=0.164 M_\odot$
and $j_0=7. 10^{52}$ g cm$^2$ s$^{-2}$)  Top: surface density
of solids in g cm$^{-2}$, as a function of distance from the star in
AU, at the times indicated. Bottom: particle radius in cm as a
function of distance from the star in AU, at the same
times. Due to limited numerical resolution the location of the
evaporation line is defined with the accuracy of $\sim 0.1$ AU. After
$1\times 10^4$ yr it stays practically constant. }
\label{solids}
\end{figure}
\begin{figure}
\resizebox{\hsize}{!}{\includegraphics{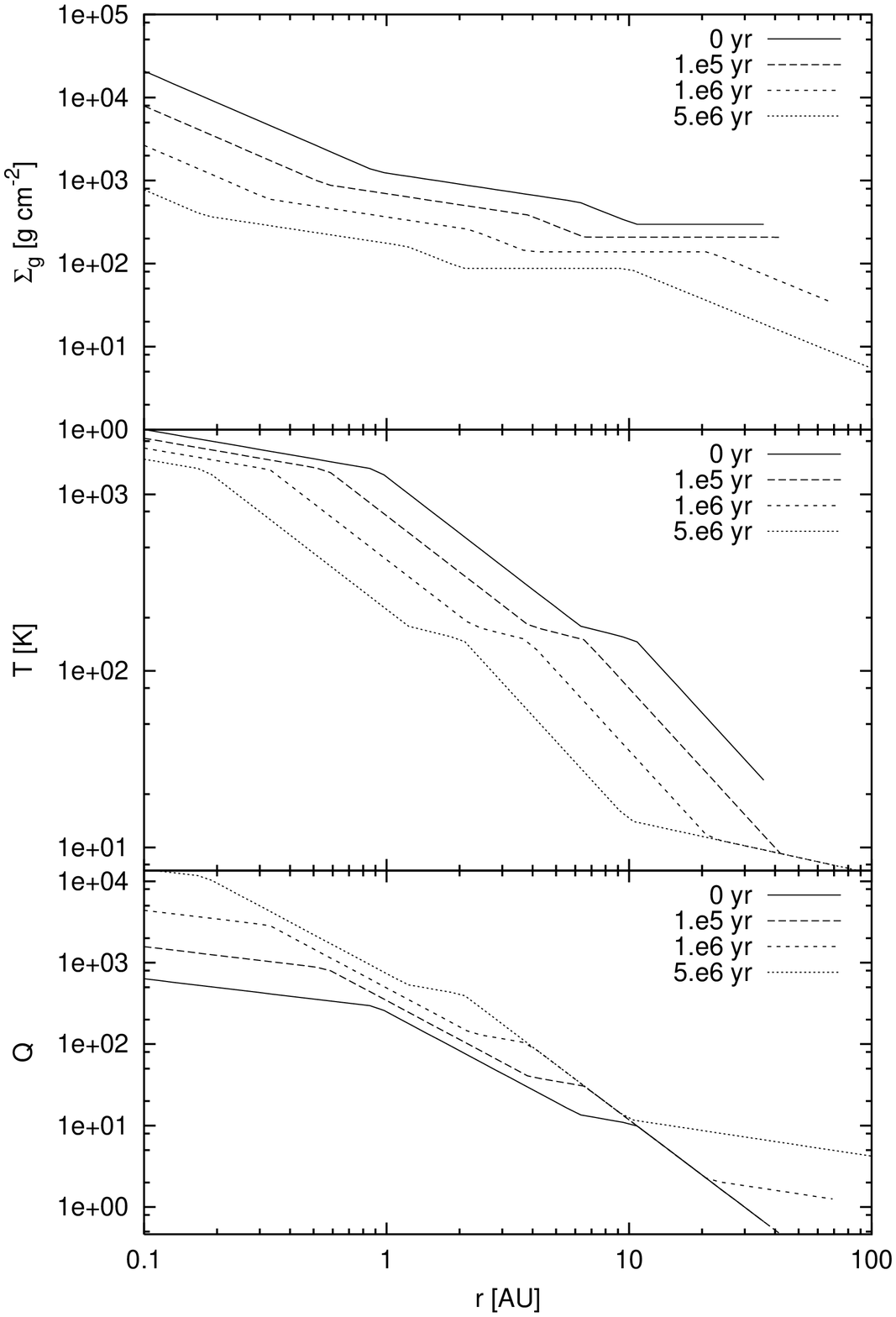}}
\caption{Evolution of the gas  component in the particular disk model
used for planet formation in the 47 UMa system. 
Top: surface density of the gas in 
g cm$^{-2}$, as a function of distance from the star in AU,
at the times indicated.
Central frame:  temperature in K as a function of distance
 from the star in AU, at the same times. Bottom: 
Toomre $Q$ parameter  as a function of distance
 from the star in AU, at the same times. }        
\label{gas}
\end{figure}

We first discuss the general results of disk evolution as a function of
$m_0$ and $j_0$. The dust and gas surface densities are evolved until
either (1) the outer edge of the dust disk falls within 0.1 AU, in which
case all the dust is assumed to have accreted onto the star, or (2)
the total elapsed time is $10^7$ yr. In the second case, it  usually
occurs   that the dust surface density distribution becomes constant
in time, well before $10^7$ yr. 

Figure~\ref{mass_rout} shows final disk properties as a function of
$m_0$ and $j_0$ for a viscosity parameter $\alpha = 10^{-2}$. The
final outer radii of the dust disk are given by the contours, and the
final solid mass is given by the greyscale. For a given $m_0$ and very
low $j_0$ all of the solid material accretes onto the star. As $j_0$
is increased, the final solid mass increases rapidly, up to a
saturation value, then remains constant with $j_0$.  For a given value
of $j_0$ the final mass of solids generally increases with $m_0$
except in the region in the ($m_0$, $j_0$) plane where accretion of
dust onto the star is significant. The grey area in
Fig.~\ref{mass_rout} indicates the region in which all particles
accrete onto the star (hereafter referred to as the dust accretion
region). The steplike shape of its upper boundary is a consequence of
the finite grid in $m_0$ (11 points) and $j_0$ (9 points). The
boundary in a good approximation corresponds to disks in which at $t =
0$ the evaporation radius is equal to the initial outer disk radius.
The final outer dust radius generally increases with 
$j_0$, and decreases with $m_0$. The region in the ($m_0$, $j_0$)
plane which is most favorable for planet formation will be discussed
below.

Figure~\ref{rin_rout}, with a similar form to Fig.~\ref{mass_rout},
shows again the outer radii of the final solid disk as contours, and
the corresponding inner radii as greyscale. The general trend is that
as the outer radius increases, the inner radius decreases.  This
effect occurs because as $m_0$ is decreased and $j_0$ is increased,
the surface density of the gas disk, and therefore its temperature,
decreases.  As a result, the evaporation radius ($R_{evap}$) is
shifted inwards.  Once the initial $R_{evap}$ is chosen, it can only
move inward during the evolution. We find that the final inner radius
of the dust disk is determined by the position of $R_{evap}$ at the
particular time when the particles just outside $R_{evap}$ become
large enough so they no longer migrate inward because of gas drag. In
initially cooler disks, $R_{evap}$ at this moment is located closer to
the star.  Note that just outside the boundary of the dust accretion
region in the ($m_0$, $j_0$) plane, the final solid disks are rather
compact, with the inner radius not much different from the outer
radius. The presence of a compact disk tends to give a solid surface
density high enough to allow the formation of a planet at relatively
small distances (inside 5 AU) from the star.

The various disk models were examined to determine which one had the
highest solid surface density at 2.1 AU at the final time. The
parameters are $m_0 = 0.164$ and $j_0 = 7$, which places the model
just above the boundary of the dust accretion region.  The properties
of the dust disk as a function of radius at different times are shown
in Fig.~\ref{solids}.  The plot shows both the solid surface density and the
particle size.  Initially the outer radius is about 40 AU and the
evaporation radius is at 0.9 AU.  The solid surface density varies from
about 10 g cm$^{-2}$ at the evaporation radius to about 2 g cm$^{-2}$
at the outer radius. Initially the particle sizes are all $10^{-3}$
cm. As a function of time, outside the evaporation radius, the
particle size increases, more slowly in the outer region because of
the lower densities and collision rates.  As the particle size
increases, the vertical thickness of the dust disk decreases.  At 100
yr some of the solid material has migrated inside the evaporation
radius, and the sharp maximum in $\Sigma_s$ actually is composed of
vapor. The region just outside the evaporation radius is depleted in
solids. At 1000 years the maximum $\Sigma_s$ in the vapor region is
higher and has been somewhat smoothed by the viscosity (in this region
the vapor is directly coupled to the gas).  The region of depletion
outside the evaporation radius is larger. At 10,000 yr the inner
maximum in the vapor region has been completely smoothed out, and
another maximum with $\Sigma_s = 25 $ g cm$^{-2}$ has formed, just outside
the evaporation radius, at 1 AU.  This region is populated by
particles that have migrated from the outer regions of the disk, but
have  grown to large enough size (20 m) so that they no longer
migrate. A short time later ($1.1 \times 10^4$ yr) this peak becomes
somewhat higher and spreads outward in radius, while the outer part of
the disk, beyond 4 AU, is strongly depleted as compared with the
initial particle density. At this time the outer radius of the solid
disk has decreased to about 10 AU.  Beyond this time the region of
high solid surface density between 1 and 3 AU does not evolve, because
the particles are large enough so they do not migrate; the particles
just increase in size. The region of high density increases somewhat
in radius as particles from the outer regions migrate into it. At $2
\times 10^5$ yr the evolution of $\Sigma_s$ stops with an inner radius
of 0.9 AU and an outer radius of 4 AU. The values of $\Sigma_s$ at 2.1
AU and 3.95 AU are 50 and $\approx 15$ g cm$^{-2}$, respectively.  The
outer value is approximate because it falls very close to the outer
edge of the solid disk.  At this time the typical particle size is 1
km; at later times the particle size would tend to increase even
further, but the model for the solid accretion is no longer valid
because it does not include gravitational effects. 
Note also that $a(r)$ approaches a constant value, consistent with the
assumption of constant planetesimal size that is made in section 2.2.

Figure~\ref{gas} shows the evolution of the gas in the particular disk
shown in Fig.~\ref{solids}. The surface density $\Sigma_g$ generally
decreases in time and the disk expands in radius, as would be expected
for a standard accretion disk. The mass of the gas decreases to 0.15,
0.11, and 0.09 M$_\odot$ at times of $1 \times 10^5,~1 \times 10^6,$
and $5 \times 10^6$ yr, respectively.  On the temperature plot, the
evaporation radius is always the first point (in radius) where the
slope changes. It moves inward from about 0.9 AU to 0.2 AU.  The slope
changes correspond to changes in the dust opacity, which is assumed to
vary as a power law in temperature, with different exponents in
different regions of temperature (Stepinski \cite{Stepinski}).  The lower portion
of Fig.~\ref{gas} shows the Toomre $Q$ stability parameter. For values above
$\approx 1$ the disk would not be expected to form planets by the
gravitational instability mechanism. The initial gas disk is in fact
gravitationally unstable outside 30 AU. The mass in the unstable
region is about 0.1 M$_\odot$, so there is at least a possibility,
untested by detailed numerical simulations, that a planet could form
rapidly by gravitational instability in the very outer region.  At
later times the disk becomes increasingly stable, and it is always
highly stable in the region from 2 to 5 AU.

Planet formation at 2.1 AU is assumed to start when the particle size
reaches 2 km, which occurs at a time of $2.1 \times 10^4$ yr (this
time corresponds to a definite upper limit of the applicability of the
disk evolution code),
This choice is somewhat arbitrary, but both the early
evolution of planetesimals and assembly of the protoplanetary core
proceed so rapidly, that their time scale is at least one 
order of magnitude shorter then the core and envelope accretion time
scale.  The $\Sigma_s$ has reached a value of 50 and does not change
in time. Assuming that there is one dominant planetary core, equation
(\ref{mdot2}) is integrated in time, using appropriate parameters for 2.1 AU,
starting with a core mass of $10^{17}$ g and ending at 1
M$_\oplus$. The calculated time is $5 \times 10^4$ yr, so the starting
time for the full planetary formation calculation is at $7 \times
10^4$ yr.  The surface boundary conditions for the forming planet are
taken from the disk model at that time, which has a mass density $\rho
= 7 \times 10^{-11}$ g cm$^{-3}$ and a temperature $T = 400$ K at 2.1
AU. At later times ($\approx 2 $ Myr) these surface values decrease to
$\rho = 4.7 \times 10^{-11}$ g cm$^{-3}$ and $T = 170$ K.

\begin{figure}
\resizebox{\hsize}{!}{\includegraphics{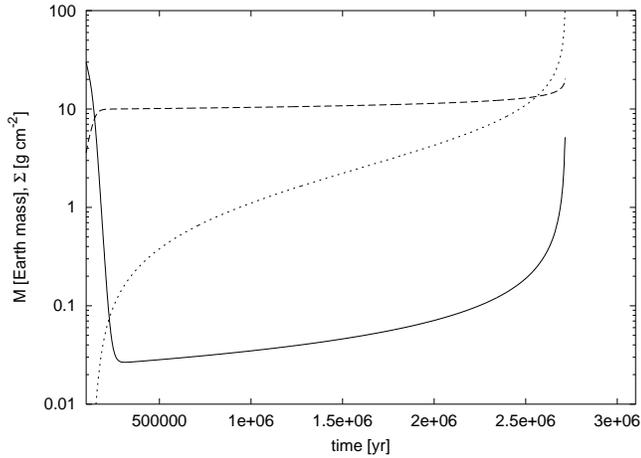}}
\caption{Formation phase of a giant planet at 2.1 AU from the
star with an assumed initial solid surface density of 50 g
cm$^{-2}$.  Dashed, dotted,  and solid lines indicate,
respectively, core mass in M$_\oplus$, envelope mass in M$_\oplus$, and
solid surface density in g cm$^{-2}$ remaining in the disk at the
location of the planet, all as a function of time in years (counted
from the beginning of the disk evolution).}
\label{in_m}
\end{figure}
\begin{figure}
\resizebox{\hsize}{!}{\includegraphics{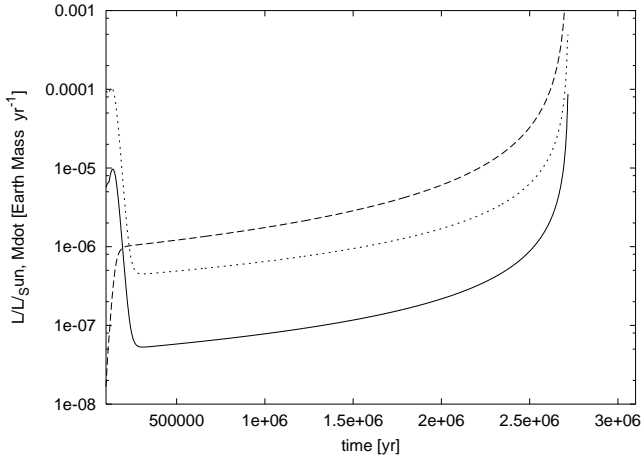}}
\caption{For the same model as in Fig.~\ref{in_m}, dotted,
dashed, and solid curves indicate, respectively, the accretion rate
of solid material onto the core in M$_\oplus$ yr$^{-1}$, the accretion
rate of gas into the envelope in M$_\oplus$ yr$^{-1}$, and the
radiated luminosity of the protoplanet, in L$_\odot$, all as a
function of time in years (counted from the beginning of the
disk evolution).}
\label{in_l}
\end{figure}

 The results are shown in Fig.~\ref{in_m} and \ref{in_l}.
Figure~\ref{in_m} shows the solid core mass, the gaseous envelope
mass, and the solid surface density in the disk as a function of time.
Figure~\ref{in_l} shows the rate of increase of core mass, the rate of
increase of envelope mass, and the radiated luminosity (in solar
units) as a function of time.  The evolution is divided into three
phases. Phase 1 is characterized by a fairly rapid increase in the
core mass during the period when the envelope mass is negligible. The
solid accretion rate peaks at $10^{-4}$ M$_\oplus$ yr$^{-1}$, then
declines as the core accretes most of the solid mass within the
feeding zone. The rapid drop in $\Sigma_s$ during this phase is
evident in Fig.~\ref{in_m}.  The core mass builds up to 10 M$_\oplus$
on a time scale of $10^5$ yr. The luminosity, which is provided by
accretion of planetesimals onto the core, peaks at about $10^{-5}$
L$_\odot$. The accretion rate of the envelope is small compared with
that of the core, but is rapidly increasing.  Phase 2 starts when the
core mass has leveled off and the accretion rates of core and envelope
become equal at $10^{-6}$ M$_\oplus$ yr$^{-1}$, which occurs at a
total elapsed time of $2.3 \times 10^5$ yr.  During this phase, which
determines the overall formation time scale, the envelope accretion
rate is a factor 2--3 larger than that of the core, so the envelope
mass builds up more rapidly than that of the core. The luminosity
remains at a low value of $10^{-7} - 10^{-6}$ L$_\odot$.  Phase 3
begins at $t = 2.6 \times 10^6$ yr when crossover mass is reached,
with envelope mass equal to core mass (13.3 M$_\oplus$ in this case),
and proceeds with a rapidly increasing accretion rate for the
envelope. The plot cuts off at 100 M$_\oplus$, but the calculation was
continued until the (minimum) mass of 47 UMa b, 2.62 M$_J$, was
reached. The formation time to final mass is 2.7 Myr, and the final
core mass is 21 M$_\oplus$.  During Phase 3, the luminosity is
provided primarily by rapid contraction of the envelope itself, and it
rapidly increases to a second peak of $\approx 10^{-2}$ L$_\odot$ (not
shown; see Bodenheimer et al. 2000).

Planet formation at 3.95 AU is assumed to start when the particle size
reaches 2 km, which occurs at a time of $4 \times 10^4$ yr.  The
$\Sigma_s$ has reached a value of $\approx 15$ and does not change in
time. The buildup time from 2 km planetesimals to a core of 1
M$_\oplus$ is calculated to be $2 \times 10^5$ yr, so the
starting time for the full planetary formation calculation is $2.4
\times 10^5$ yr.  The surface boundary conditions for the forming
planet are taken from the disk model at that time, which has a mass
density $\rho = 2 \times 10^{-11}$ g cm$^{-3}$ and a temperature $T =
170$ K at 3.95 AU. At $2 \times 10^6$ yr these values have decreased
to $\rho = 1.6 \times 10^{-11}$ g cm$^{-3}$ and $T = 111$ K.

\begin{figure}
\resizebox{\hsize}{!}{\includegraphics{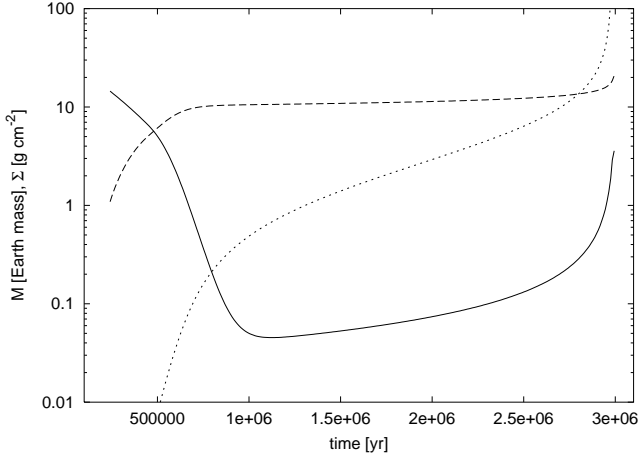}}
\caption{Formation phase of a giant planet at 3.95 AU from the
star with an assumed initial solid surface density of 15 g
cm$^{-2}$.  Dashed, dotted,  and solid lines indicate,
respectively, core mass in M$_\oplus$, envelope mass in M$_\oplus$, and
solid surface density in g cm$^{-2}$ remaining in the disk at the
location of the planet, all as a function of time in years (counted
from the beginning of the disk evolution).}
\label{out_m}
\end{figure}
\begin{figure}
\resizebox{\hsize}{!}{\includegraphics{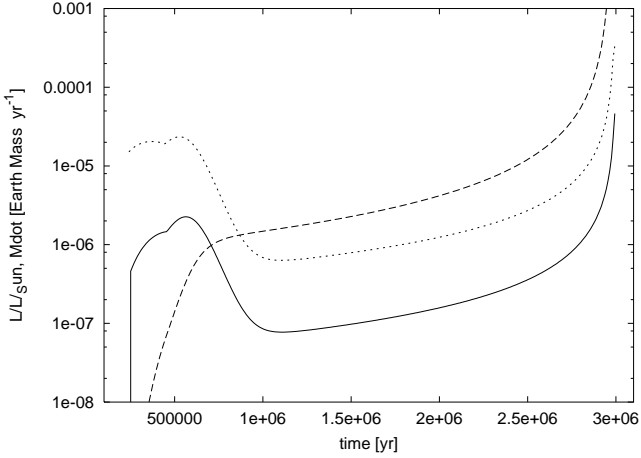}}
\caption{For the same model as in Fig.~\ref{out_m}, dotted,
dashed, and solid curves indicate, respectively, the accretion rate of
solid material onto the core in M$_\oplus$ yr$^{-1}$, the accretion
rate of gas into the envelope in M$_\oplus$ yr$^{-1}$,  and the radiated
luminosity of the protoplanet, in L$_\odot$, all as a function of time
in years (counted
from the beginning of the disk evolution).}
\label{out_l}
\end{figure}

The planet at 3.95 AU is assumed to be formed independently of the one
at 2.1 AU; in fact their feeding zones do not overlap. This 
assumption is somewhat restrictive as in principle another
protoplanetary core(s) could form between 2.1 and 3.95 AU. However we
just want to demonstrate that the \textit{in situ} formation of the 47
UMa system is possible and therefore we do not consider any other
evolutionary scenarios. The results are shown in Fig.~\ref{out_m} and
\ref{out_l}, which show the same quantities as in Fig.~\ref{in_m} and
\ref{in_l}, respectively.  During Phase 1, the core accretion rate
increases to a maximum of $2.5 \times 10^{-4}$ M$_\oplus$ yr$^{-1}$
and then declines as $\Sigma_s$ in the disk is depleted.  The core
mass builds up to 10 M$_\oplus$ on a time scale of $3 \times 10^{5}$
yr.  The luminosity during Phase 1 peaks at about $2.5 \times 10^{-6}$
L$_\odot$, somewhat lower than that for the planet at 2.1 AU.  Phase 1
ends when the accretion rate of the envelope, previously low, equals
that of the core, at $8.5 \times 10^{5}$ yr, a factor 3.7 longer than
the corresponding time for the planet at 2.1 AU.  Phase 2, however, is
very similar in the two cases with regard to time scale, accretion
rates, and luminosity. The crossover mass of 13 M$_\oplus$ is reached
at $2.8 \times 10^{6}$ yr, slightly later than that for the planet at
2.1 AU.  The calculation was continued until the minimum mass of 47
UMa c, 0.89 M$_J$, was reached after a time of $3.0 \times 10^6$ yr,
only $3 \times 10^5$ yr longer than that for the inner planet. The
final core mass has increased slightly since crossover to 16
M$_\oplus$.

\section{Conclusions }
\begin{figure}
\resizebox{\hsize}{!}{\includegraphics{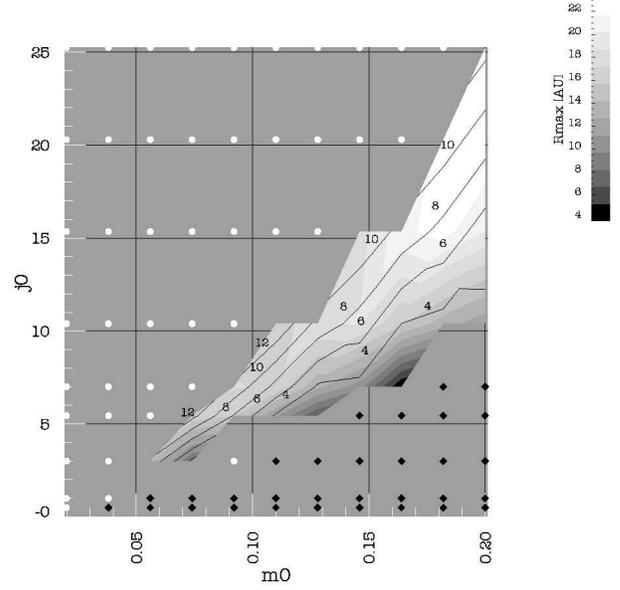}}
\caption{ As functions of the initial disk mass $m_0$ (in solar
masses) and angular momentum $j_0$ in units of $10^{52}$ g
cm$^2$ s$^{-1}$, the contours and grey scale,
respectively, give the inner and outer radius of the region around the
central star where giant planet formation is possible in a maximum of
3 Myr.  White circles indicate disk models where the solid surface
density is everywhere below the critical value for planet formation.
Diamonds indicate disks in which all of the solid material accretes
onto the star. The disk viscosity parameter $\alpha = 1 \times
10^{-2}$.}
\label{crit_2}
\end{figure}
\begin{figure}
\resizebox{\hsize}{!}{\includegraphics{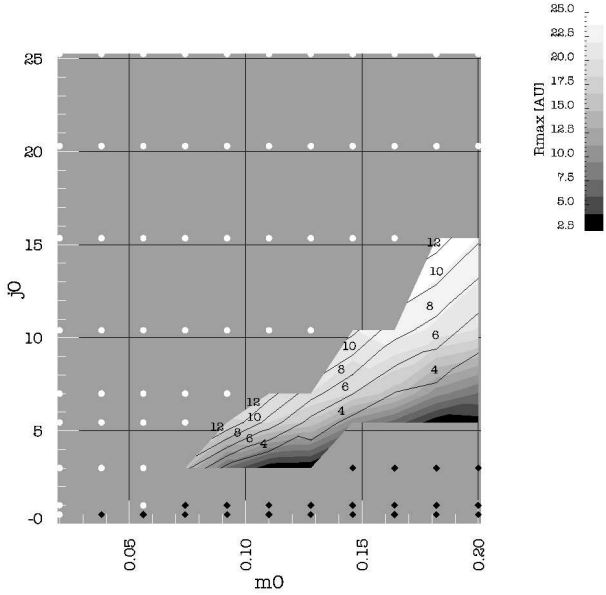}}
\caption{The symbols and grey scale have the same meaning as in Fig.~\ref{crit_2}, 
for $\alpha = 1 \times 10^{-3}$.}
\label{crit_3}
\end{figure}

These numerical results allow us to reach the following conclusions:
(1) There exists a disk model which allows the formation of both of
the planets in the 47 UMa system in about 3 Myr at their present
distances from the star. (2) The initial disk may be significantly
less massive than the one required by Bodenheimer et al.  (2000). At
2.1 AU their gas density had to be as high as $2.1 \times 10^4$ g
cm$^{-2}$, while the disk used in the present calculation had
$\Sigma_g = 10^{3}$ g cm$^{-2}$ at 2.1 AU at the beginning of the disk
evolution. (3) The solid cores of both planets are relatively small,
21 and 16 M$_\oplus$ for the inner and outer planet, respectively. In
comparison, the core mass for the inner planet in model U2 of
Bodenheimer et al. (2000), which formed in about 2 Myr, was 69
M$_\oplus$.  (4) The planet at 2.1 AU formed in a much shorter
time (2.7 Myr vs. 18.6 Myr) than that with the same assumed solid
surface density in the the calculations of Bodenheimer et al. (2000b).
The main reason is that the grain opacity in the present calculation
is up to a factor 100 lower than that assumed by Bodenheimer et
al.(2000b).  That effect is known to lead to shorter formation times.
A test calculation for the inner planet was made in which the grain
opacity was reset to interstellar values, as used by Bodenheimer et
al. (2000b), all other effects remaining the same. The formation time
turned out to be $1.2 \times 10^7$ yr, more than a factor 4 longer
than the 2.7 Myr obtained with the reduced opacity, but about 30\%
shorter than in Bodenheimer et. al (\cite{BHL}). The latter difference
is explained by the fact that the earlier calculation had different
surface boundary conditions, a different procedure for calculating
$\dot M_Z$, and a higher core density (5 g cm$^{-3}$). (5) In the
presence of the fully formed planet at 2.1 AU, there is still some
solid material left in the disk between 1 and 1.5 AU, with $\Sigma_s
\approx 40$ g cm$^{-2}$ and a mass of about 6 M$_\oplus$. Formation of
another giant planet in this region is not possible; a numerical
calculation shows that the formation time would be much longer than
the lifetime of the gas disk. The minimum solid surface
density required to form Jupiter size planet in this region in 3 Myr
is $\sim 100$ g cm$^{-2}$ (see below).

These results lead to the question: in what kinds of disks is it
possible to form Jupiter size planets on a 3 Myr time scale?
We calculated approximate planet formation models, fitted to the
results of Section 3, to determine, at various distances from a 1
M$_\odot$ star, the minimum solid surface density $\Sigma_{s,min} $
needed to form a giant planet in 3 Myr.  The results for
$\Sigma_{s,min} $ range from 100 g cm$^{-2}$ at 1.0 AU to 9 g
cm$^{-2}$ at 5 AU to a minimum of 3 g cm$^{-2}$ at 15 AU. Then they
increase slowly outward to 4.6 g cm$^{-2}$ at 50 AU.  
This result can be explained as follows. At small $R$ the protoplanet spends
most of the evolutionary time in phase 2. The time scale for phase 2 depends
strongly on the core mass (see Pollack et al. 1996 for a detailed explanation),
so that in order to form the planet in a given time, say 3 Myr, the core mass
must exceed some critical mass $M_{crit} \sim 15 M_{\oplus}$.
 Thus the necessary condition to form  the planet at a given location is
that the isolation mass, $M_{iso}$ is not smaller then $M_{crit}$. According to Pollack
et al. (1996)
\begin{equation}
   M_{iso}=C_1 (R^2 \Sigma_s)^{3/2}
\label{miso}
\end{equation}
Setting $M_{iso}=M_{crit}=\mathrm{const}$ we recover the rapid increase of
$\Sigma_{s,min}$  toward the center of the disk as observed in numerical
results.On the other hand, at large $R$ the evolutionary time scale is determined by
the length of phase 1 in which the core is assembled. Here, the necessary
condition for the planet to form within a prescribed time is that the mass
of the core reaches $M_{crit}$. Setting
$M_{crit}=\mathrm{const}$ and employing formula (\ref{mdot2}) we get \[
\Sigma_{s,min} \sim R^{1/2}, \] again in a rough agreement with the
numerical results.

We then examined
all the disk models shown in Fig.~\ref{mass_rout} and \ref{rin_rout}
to find the range of distances $R_{min} \leq R \leq R_{max}$ at which
the final $\Sigma_s$ exceeds the local $\Sigma_{s,min}$.
Figure~\ref{crit_2} shows the results for disk models with $\alpha =
10^{-2}$, as a function of the fundamental disk parameters $m_0$ and
$j_0$. The grey scale gives the values of $R_{max}$ and the contours
give the values of $R_{min}$.  Note that the region of possible planet
formation at less than 5 AU is very limited.  In fact we can estimate,
consistent with this model, that the minimum radius at which giant
planet formation is possible is about 2 AU.  Also, the maximum radius
is estimated at 23 AU. In general, if $R_{min}$ is small, $R_{max}$ is
also relatively small.  In models marked by a white circle, there is
solid material present at the end of the evolution, but $\Sigma_s$ is
everywhere too low to form planets. In models marked by a diamond, all
of the solid material accretes onto the star. Note that the disk
models do not include ice grains, which will be included in future
calculations. However the minimum radius for planet formation is not
expected to change if ice is included.

Figure~\ref{crit_3} shows the same results as Fig.~\ref{crit_2} except
for a disk viscosity parameter $\alpha = 10^{-3}$. The minimum radius
for planet formation in this case is somewhat smaller, about 1 AU, and
the maximum is somewhat larger, about 25 AU. The main effect of the
reduction in $\alpha$ is a more extended region in which the dust
survives (Stepinski \& Valageas \cite{SV97}). The inner radii of the final
dust disk are smaller because the lower-$\alpha$ disks have lower
temperatures as a result of a smaller energy generation rate by
viscosity at a given surface density. Thus the evaporation radius
tends to be at smaller distances.  On the other hand, the outer radii
of the dust disks are larger because the inward drift of particles is
slower.  Another effect of reducing $\alpha$ is to reduce the maximum
$j_0$ for which planet formation is possible.  Generally as $j_0$ is
increased for a given $m_0$, the gas and dust disks become more
extended and therefore have lower average surface density.  For the
same disk parameters except for a lower $\alpha$, the dust disk is
more extended, so the $\Sigma_s$ falls everywhere below the critical
value at smaller $j_0$.

In summary, the planets in the 47 UMa system can be explained by the
core accretion -- gas capture process in a disk with reasonable
parameters (mass 0.16 M$_\odot$ and outer initial radius 40 AU) in
which the dust component consists of high-temperature silicates.  The
inclusion of ice grains in the model would probably make little
difference in the planet formation process: a corresponding disk model
calculated with only ice grains initially present resulted in the
accretion of all of the ice onto the star.  The planets can reach
their observed minimum masses in 3 Myr.  The main factor enabling  
their formation is that the surface density of solids in
the region 2--4 AU is considerably higher than in the `minimum mass'
solar nebula.

 Migration of the planets as a result of gravitational interactions
with the disk was not included.  However migration is still a
possibility since with slightly changed disk parameters  the
planets might be formed at much larger distances from the star (note
that  in the
present model the inner planet is located 
at about the minimum distance from the star where it is
possible to form a giant planet in an $\alpha = 10^{-2}$ disk).

Future improvements of the model should include (1) interactions
between different types of solid particles, (2) particle models in
which a range of particle sizes at each position in the disk is
considered, (3) improvements in the dust opacity in the envelopes of
the protoplanets, which influences the time scale of Phase 2,  and
(4) a  better description of the boundary between the envelope of the
planet and the disk; in particular, the presence of the secondary
(circumplanetary) disk described by Ciecielag  et al. (\cite{Ciecielag}) and
D'Angelo  et al.  (\cite{DAngelo02}). Finally, the effect of increased or reduced metal
abundance on the region of parameter space in which planet formation
is possible should also be investigated.

\begin{acknowledgements}
This work was supported in part through NASA grants NAG5--9661 and
NAG5--9526 from the Origins of Solar Systems Program and through
Polish Committee for Scientific Research grant 2P03D01419.  We thank
Debra Fischer for providing updated observational data on the
properties of the 47 UMa system.
\end{acknowledgements}

\end{document}